%% file: pritychenko20.tex
\documentclass{aastex62}

\shorttitle{
Determination of Solar System R-Process ...
}
\shortauthors{ B. Pritychenko}

\begin{document}

\title{
Determination of Solar System R-Process Abundances using ENDF/B-VIII.0 and TENDL-2015 libraries
}

\correspondingauthor{Boris Pritychenko}
\email{pritychenko@bnl.gov}

\author{Boris Pritychenko}
\affiliation{National Nuclear Data Center, Brookhaven National Laboratory, \\ Upton, NY 11973-5000, USA}



\begin{abstract}

Recent multi-messenger detection of the  binary neutron star merger (GW170817) energized the 
astrophysical community and encouraged further research for determination of nuclear physics observables. 
Comprehensive studies of atomic nuclei in the cosmos provide  an opportunity for investigating these astrophysical phenomena 
and acquiring complementary information on stellar nucleosynthesis processes 
that can be verified using the latest nuclear data. 

Evaluated Nuclear Data File (ENDF) libraries contain complete collections of reaction cross sections over the energy range relevant to astrophysics, 
 fission yields and decay data. These data collections have been used 
worldwide in nuclear science, industry and national security applications. 
There is great interest in exploring the  ENDF/B-VIII.0  and TALYS Evaluated Nuclear Data Library (TENDL-2015)  
for nuclear astrophysics purposes and comparing findings with the Karlsruhe Astrophysical Database of Nucleosynthesis in Stars (KADoNiS). 

The Maxwellian-averaged cross sections (MACS) and astrophysical reaction rates have been calculated using 
the ENDF/B-VIII.0 and TENDL-2015 evaluated data sets. The calculated cross sections were combined with the solar system 
abundances and fitted using the classical model of stellar nucleosynthesis. Astrophysical rapid- and slow-neutron capture, $r$- and $s$-process, respectively, 
abundances were obtained from present data and compared with  available values.  
Further analysis of MACS reveals potential evaluated libraries data deficiencies and a strong need for 
new measurements. The current results demonstrate a large nuclear astrophysics potential of evaluated libraries 
and mutually beneficial relations between nuclear industry and research efforts.
\end{abstract}



\section{Introduction} \label{sec:intro}

The indirect findings of $r$-process production of gold, platinum, and  lanthanide elements  in neutron stars merger \citep{17Pia,17Sma,17Tan} 
renewed interest in stellar nucleosynthesis abundance calculations. These calculations strongly 
rely on the quality of underlying nuclear data  and astrophysical models  \citep{88Rol}. ENDF is a primary nuclear reaction data library for applications, and  it is focused on nuclear materials (target nuclei) near the valley of stability. Although evaluated libraries can only reliably be used to model $s$-process (due to limitation in isotope inventory off-stability), one can apply these data to estimate $r$-process contributions along the $s$-process path. The release of ENDF/B-VIII.0 evaluated 
nuclear data library by the Cross Section Evaluation Working Group (CSEWG) collaboration \citep{18Bro} creates a unique opportunity to verify  nucleosynthesis results, using the latest evaluated neutron cross sections.

In the current paper, Maxwellian-averaged cross sections  and astrophysical reaction rates at $kT$=30 keV are calculated
from ENDF/B-VIII.0 and TENDL-2015 \citep{15Kon} libraries using the techniques described in Ref. \citep{12Pri}. 
The computed values are further analyzed and compared  with available data,  and multiple recommendations for potential applications and data quality 
improvements have been produced.  

Finally, ENDF cross sections are used to calculate $s$- and deduce $r$-process solar system abundances using the classical model of stellar nucleosynthesis \citep{74Cla}. The deduced $r$-process abundances provide a basis for verification of the previous results \citep{93Pal,07Arn}, comparison with astrophysical observations of chemical composition 
of the $r$-process dwarf galaxies \citep{18Fre}, and, perhaps, future measurements of elements and isotopes production in neutron star mergers.

\section{Nuclear Data Libraries} \label{sec:lib}

Nuclear data program include nuclear reaction, structure and decay libraries. These libraries contents incorporate bibliographical data that are compiled in the Nuclear Science References (NSR) database \citep{11Pri}. 
\subsection{Nuclear Reaction Libraries}
First nuclear reaction physics data compilations have performed during the Manhattan Project  \citep{47Chicago}, and this work has been continued on a permanent basis at the Brookhaven National Laboratory (BNL) since the early 1950s \citep{52Hughes}. The BNL group summarized its findings in the now famous BNL-325 report  \citep{BNL325,06Atlas}, and these efforts laid a foundation for worldwide nuclear data compilation and evaluation activities in subsequent years.  Over the years compilations have matured and evolved into unique nuclear physics database Sigma Center Information Storage and Retrieval System (SCISRS) database that is presently known as EXFOR (Experimental Nuclear Reaction Data) \citep{14Otu,18Zer}. The EXFOR library \citep{14Otu} and its computer database, https://www.nndc.bnl.gov/exfor  \citep{18Zer}, are primary source of experimental  neutron, charged-particle, and photonuclear  nuclear reaction data compilations. Several nuclear reaction model codes such as  EMPIRE and TALYS  \citep{2007Empire,2012TALYS} include EXFOR data. 

Evaluated Nuclear Data File (ENDF)  is a major nuclear reaction library. This library includes evaluated (recommended) cross sections, spectra, angular distributions, fission product yields, thermal neutron scattering, photo-atomic, and decay data, with an emphasis on neutron-induced reactions. Its evaluations are based on nuclear model predictions and normalized to experimental data, with an exception of neutron resonance region, where priority is directly given to experimental data.  
ENDF  library contains multiple sublibraries and complete data collections for a particular isotope or element (ENDF material). These data collections cover all neutron reaction channels within the 10$^{-5}$ eV - 20 MeV energy range. In several evaluations, this energy range is extended beyond the 20 MeV upper limit. The ENDF data are stored in the internationally-adopted ENDF-6 format that is maintained by the CSEWG collaboration, and they are publicly available at the National Nuclear Data Center (NNDC) website, https://www.nndc.bnl.gov/endf. 
The neutron sublibrary is a basis of ENDF, and its data are used in many applications, nuclear power plant design, criticality safety, shielding, and national security,  GEANT, and MCNP computer codes simulations \citep{GEANT,MCNP}. ENDF libraries are  very popular worldwide, and neutron cross section evaluations are conducted internationally for ENDF,  JEFF (Joint Evaluated Fission and Fusion) File, TENDL,  JENDL (Japanese Evaluated Nuclear Data Library), ROSFOND (ROSsijsky  File  Otsenennykh  Neutronnykh Dannykh) and CENDL (Chinese Evaluated Nuclear Data Library)  \citep{18Bro,11Kon,15Kon,11Shi,07Zab,11Ge} in the United States, Western Europe, Japan, Russian Federation, and China, respectively. In complement with traditional ENDF/B libraries, TENDL-2015 or Talys Evaluated Nuclear Data Library~\citep{15Kon}, is mostly calculated using the TALYS nuclear model code~\citep{2012TALYS}. The TALYS code calculations produce the most extensive and complete collection of target materials and include all reaction channels and their uncertainties.

For many years nuclear astrophysicists concentrated on their own  data sets such as the Karlsruhe Astrophysical Database of Nucleosynthesis in Stars (KADoNiS) library ~\citep{06Dil,14Dil}, REACLIB ~\citep{00Rau,01Rau,10Cyb}, BRUSLIB \citep{13Xu} and STARLIB \citep{13Sal}    collections of reaction rates  were  used in nuclear astrophysics calculations, while the ENDF libraries were largely unexplored. The calculation of JENDL-3.3 library MACS and reaction rates \citep{05Nak} demonstrated an applicability of evaluated libraries for astrophysical tasks and, in some cases, the calculated values  have deviated from the KADoNiS values. These cross section discrepancies have been resolved in subsequent BNL works  \citep{10Pri,12Pri,18Bro} when MACS, reaction rates, and their uncertainties for multiple libraries have been produced and analyzed. The ENDF evaluated cross sections and reaction rates could provide essential ingredients for  stellar nucleosynthesis models, and they have not been previously examined.   Therefore, the applicability of evaluated neutron cross sections for astrophysical modeling and $s$-process nucleosynthesis calculations will be investigated in the present work.

\subsection{Nuclear Structure and Decay Libraries}

The primary source for recommended structure and decay data worldwide is the Evaluated Nuclear Structure Data File (ENSDF) database \citep{ensdf}. This project originates from Oak Ridge National Laboratory, and it is currently run by the National Nuclear Data Center (NNDC), BNL on the behalf of the U.S. Nuclear Data program. ENSDF evaluations are based on experimental data, and they are periodically updated on a 8-12 years basis. To improve the currency of ENSDF the eXperimental Unevaluated Nuclear Data List (XUNDL) database  \citep{XUNDL} was launched more than 20 years ago at McMaster University, Canada. XUNDL is an extensive collection of nuclear structure and decay data compilations in support of the Evaluated Nuclear Structure Data File (ENSDF) database \citep{ensdf}, and both databases share the common format. ENSDF and XUNDL data reflect the current state of development of nuclear science. 

Historically, nuclear physicists first studied properties of stable and long-lived nuclei and later moved to short-lived species. During these times many experimental techniques were at the earlier stages of developments and detectors were rather simple. As the result of that quite a few findings for the stable and long-lived nuclei were not very precise. In the present days nuclear science is focused on properties of extremely neutron- and proton-rich nuclei, and scientists mostly ignore nuclei near the valley of stability. This creates a very strong limitation on $s$-process branchings computations because of the lack of precise $\beta$-decay half lives. In fact, imprecise values for $\beta$-decay rates would severely distort such branching calculations, and render them as not very reliable. Therefore, it is necessary to use the present-day nuclear physics detector systems and revisit the properties of nuclei near the valley of stability. Unfortunately, a very few experimental groups are involved in such activities, and researchers are mostly exploring nuclei near the neutron and proton driplines.

\section{Stellar Nucleosynthesis Calculations}  \label{sec:snc}

There are multiple nuclear astrophysics processes that are responsible for the present variety of elements and isotopes in the Universe \citep{88Rol};    
the slow ($s$) and rapid ($r$)  neutron capture processes are responsible for production of medium and heavy nuclei. 
The $s$-process occurs in Asymptotic Giant Branch (AGB)  and Red Giant stars, while  the $r$-process may take place in neutron star mergers.
 Further analysis of the nuclear chart helps to identify pure $s$-, $r$-process, and mixed ($s$ + $r$) nuclei.  
 Using $s$-process modeling, the author will calculate the slow neutron capture isotope abundances, subtract them from presently-observed values, and attribute the 
 remainders to  $r$-process isotopes production.

\begin{figure}
\begin{center}
\includegraphics[width=0.75\textwidth]{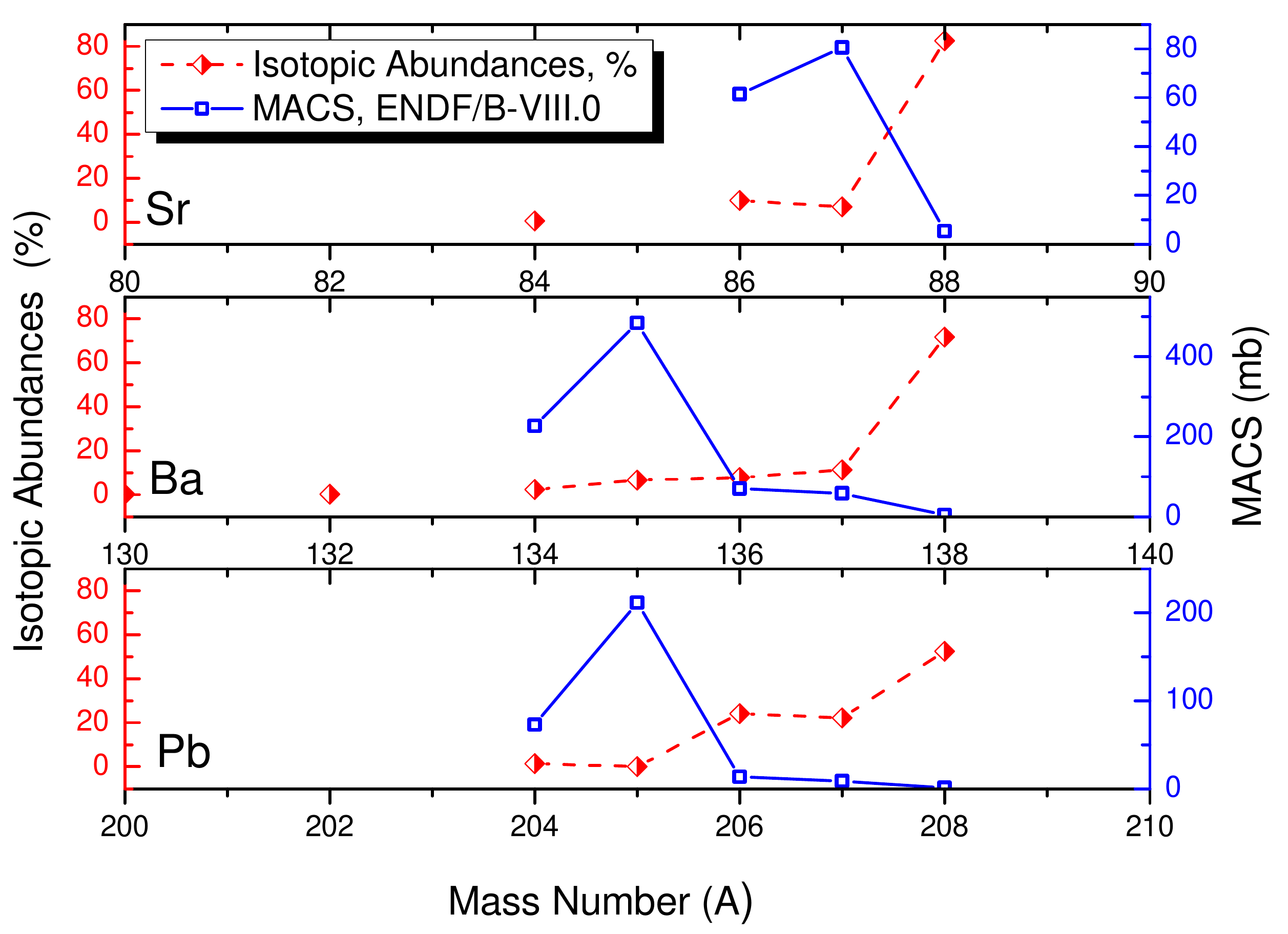}
\end{center}
\caption{Sr, Ba, Pb isotopic abundances  and ENDF/B-VIII.0 Maxwellian-averaged cross sections  at $kT$=30 keV \citep{18Bro}. \label{fig:magic}}
\end{figure}

The $s$-process is the fundamental nuclear physics mechanism that is responsible 
for production of $\sim50 \%$ of stable nuclei beyond iron, and its rate is governed by neutron properties.  
To illustrate the impact of magic neutron numbers, one should consider the isotopic abundances of strontium, barium, and lead nuclei in conjunction with MACS. The data shown in Fig. \ref{fig:magic} demonstrate a strong anti-correlation  between isotope abundances and Maxwellian-averaged neutron cross sections and the important role of closed neutron shells in $^{88}$Sr, $^{138}$Ba, and $^{208}$Pb.
Further analysis of these and other related data, leads to the  three $s$-process Golden Rules \citep{16Lug}
\begin{itemize}
\item Neutron closed shell (magic) nuclei have low neutron-capture cross sections, and they act as bottlenecks.
\item Between neutron bottlenecks, the abundances are in equilibrium.
\item Branching points may occur on the $s$-process path when a $\beta$-decay rate competes with neutron  capture.
\end{itemize}
Further analysis of the nuclear chart shows that the $s$-process proceeds along the nuclear valley of stability matching the ENDF materials list, and its path is defined by neutron capture cross sections and flux distributions that are outlined in the following subsection. 

\subsection{Maxwellian-averaged Cross Sections and Astrophysical Reaction Rates}

The incident neutrons quickly thermalize in the star interior and follow  the Maxwellian spectrum. Maxwellian-averaged cross sections (MACS) are described as   

\begin{equation}
\label{myeq.max3}
\sigma^{Maxw}(kT) = \frac{2}{\sqrt{\pi}} \frac{a^{2}}{(kT)^{2}}  \int_{0}^{\infty} \sigma(E^{L}_{n})E^{L}_{n} e^{- \frac{aE^{L}_{n}}{kT}} dE^{L}_{n},
\end{equation}
where $a = m_2/(m_1 + m_2)$, {\it k} and {\it T} are the Boltzmann constant and temperature of the system, respectively,  and $E$ is an energy of relative motion of 
the neutron with respect to the target \citep{88Rol,05Nak}. $E^{L}_{n}$ is a neutron energy in the laboratory system,  while $m_{1}$ and $m_{2}$ are masses of 
the neutron and the target nucleus, respectively.

The astrophysical reaction rate, $R$, is defined as $R$ = $N_{A}$$\langle \sigma \upsilon \rangle$, where $N_{A}$ is the Avogadro number. To express reaction rates in [$cm^{3}$/mole s] units, an additional factor of $10^{-24}$ is introduced
\begin{equation}
\label{myeq.rrates}
R(T_{9}) = 10^{-24}N_{A}\sigma^{Maxw}(kT)\upsilon_{T},
\end{equation}
where $\upsilon_{T}$ is in units of [cm/s] and temperature, $kT$, in units of energy ({\it e.g.} MeV) is related to that in Kelvin ({\it e.g.} 10$^{9}$ K) as $T_{9}$=11.6045 $kT$.

These two equations have been used in the present work to  calculate ENDF/B-VIII.0 and TENDL-2015 MACS and reaction rates  
 within the typical range of energies.  
Prior to calculations the  neutron resonance region evaluated data had been Doppler reconstructed at T=293.16 K with the PREPRO code \citep{Prepro}.  
 The numerical values of ENDF and TENDL Maxwellian-averaged cross sections and reaction rates at $kT$=30 keV 
are given in  appendix Table \ref {tableCSRR} as supplementary data. These tabulated values are used in the subsequent subsections as nuclear data inputs.

\subsection{Classical Stellar Nucleosynthesis Model}

The recommended nuclear industry data have not been explored in astrophysics. Therefore, a relatively simple, reliable and well understood classical model of stellar nucleosynthesis was chosen. The model  transparency is of paramount importance for  testing of  new data types. In addition, its empirical systematics for heavy nuclei that are not affected by branchings, are reproduced with a mean square deviation of only 3$\%$ \citep{11Kap}. The classical model is based on a phenomenological and site-independent approach, and it assumes that the seeds for neutron captures are made entirely of $^{56}$Fe.  
The $s$-process abundance of an isotope $N_{(A)}$ depends on its precursor $N_{(A-1)}$ quantity as in
\begin{equation}\label{eq:class1}
\frac{dN_{(A)}}{dt} = \lambda_{n (A-1)} N_{(A-1)} - \big[ \lambda_{n (A)} + \lambda_{\beta (A)} \big] N_{(A)},
\end{equation}
where $\lambda_{n}$ is a neutron capture rate, and $\lambda_{\beta} = \frac{ln 2}{T_{1/2}}$ is $\beta$-decay rate for radioactive nuclei. 
Assuming that the temperature and neutron density are constant,  and ignoring  $s$-process branchings, the previous formula simplifies to 
\begin{equation}\label{eq:class2}
\frac{dN_{(A)}}{dt} = \sigma_{(A-1)} N_{(A-1)} - \sigma_{(A)} N_{(A)}.
\end{equation}

Equation  \ref{eq:class2} was solved analytically  for an exponential average flow of neutron exposure 
assuming that  temperature remains constant over the whole time scale of the $s$-process \citep{74Cla}. The product of MACS and isotopic abundance ($\sigma_{(A)} N_{(A)}$) was written as
\begin{equation}\label{eq:class3}
\sigma_{(A)} N_{(A)} = \frac{f N_{56}}{\tau_{0}} \prod_{i=56}^{A} \big[ 1+ \frac{1}{\sigma(i) \tau_{0}} \big]^{-1},
\end{equation}
where  $f$ and $\tau_{0}$ are neutron fluence distribution parameters, and $N_{56}$ is the initial abundance of $^{56}$Fe seed. 
Finally, at the $s$-process equilibrium, the equation \ref{eq:class2} becomes
\begin{equation}\label{eq:class4}
\sigma_{(A-1)} N_{(A-1)} = \sigma_{(A)} N_{(A)} = constant.
\end{equation}

Further research shows that the solar system $s$-process abundances originate from a superposition of the
two major exponential distributions of time-integrated neutron exposure: weak
component (responsible for the production of 70 $\leq$ A $\leq$ 90 nuclei), and the main component (for 90 $\leq$ A $\leq$ 204 nuclei). 
It is difficult to predict the abundances of lead isotopes beyond $^{204}$Pb
because the abundance of $^{206}$Pb is altered by an $\alpha$-decay of the  $s$-process end-product  $^{210}$Po, and 
  $^{206,207}$Pb and $^{208}$Pb  abundances include daughter products of   $^{238,235}$U and  $^{232}$Th radioactive chains, respectively.

The classical model and its results have been widely used over the years  \citep{74Cla,82Kap,94Mey,11Kap}, 
and it provides an excellent description of $s$-process abundances in the solar system for heavy nuclei. In recent years, several improvements   
in astrophysical modeling and nuclear data have been made.  S. Goriely has developed a multi-event $s$-process (MES) model \citep{99Gor} to address issues with a presumed exponential form  of the neutron exposures and detailed evaluation of uncertainties, and  the MES model was eventually enhanced to describe $r$-process abundances \citep{07Arn}. Using  the latest data  for $^{142}$Nd cross sections, the preference of a stellar model over the classical has been suggested by others \citep{99Arl}.   At the same time, recent nuclear data analysis showed many cases  when the commonly-accepted nuclear data sets are not accurate \citep{15Pri}, including 5-7$\%$ deficiency in the KADoNiS  0.3 gold standard \citep{15Car,18Car,16Moh}. These developments triggered  KADoNiS library update that is currently underway \citep{18Rei}. The last result provides an alternative explanation of $^{142}$Nd cross section modeling issues.   In conclusion, the present discussion highlights the challenges in nuclear astrophysics calculations, importance of high quality input data, and value of well-understood and reproducible results. Therefore, the author would use the well-tested classical model, that is not impacted by the previously discussed $\beta$-decay deficiencies, in order to evaluate the astrophysical potential  of nuclear data libraries, produce the corresponding $s$- and $r$-process abundances, and document potential issues with the reaction data.

\subsection{Stellar Enhancement Factors (SEF)}

Presently discussed evaluated and nuclear astrophysics libraries are based room temperature measurements while stars temperatures vary within 0.01-10 GK range.  Stellar enhancement factors affect cross section and astrophysical reaction rate values in plasma environment. The SEF factors originate from stellar reaction cross sections  $\sigma^{*}$ that could be estimated as a sum of the cross sections $\sigma_{x}$ for the excited states $x$ with excitation energy $E_{x}$ and spin $J_{x}$, weighted with the Boltzmann excitation probability \citep{10Pan}
\begin{equation}
\label{myeq.sef}
\sigma^{*} = \frac{\Sigma_{x} (2 J_{x} +1) \sigma^{x} e^{\frac{-E_{x}}{kT}}}{\Sigma_{x} (2 J_{x} +1) e^{\frac{-E_{x}}{kT}}}.
\end{equation}

Calculations of SEF rely on an availability of high-quality nuclear data for excitation energies  and spins across the nuclear chart. Level energies near the valley of stability  are relatively known while spin and parity assignment is one of the most difficult tasks in nuclear physics. Incomplete nuclear structure information for many nuclei introduces certain limitations on the factors applicability at the present state of development of nuclear physics. Several proposals to verify SEF experimentally have been discussed \citep{02Koe,09Boy}; nonetheless, the required conditions have not been achieved at the National Ignition Facility (NIF) at Lawrence Livermore National Laboratory. In the mean time, the  theoretical factors of Rauscher {\it et al.} from the KADoNiS website \citep{09Rau,98Rau,98Ra} were used for $s$-process calculations in the present work.

\subsection{Neutron fluence of s-process nuclei}

Previously,  $s$-process data have been analyzed and fitted from $^{56}$Fe to $^{210}$Po  as a sum of  the weak and main  components described by  the Eq. \ref{eq:class3} of Ref. \citep{82Kap}. 
In the fit of a weak $s$-process component, F. K{\"a}ppeler {\it et al.} have used the contemporary values of abundances \citep{81Cam} and cross sections \citep{82Kap}. In addition, they included $^{88}$Sr, $^{89}$Y, and $^{90}$Zr 
to overcome a relatively small number of $s$-process only nuclei in the $A<90$ region. These authors have argued that the above-mentioned nuclei solar system abundances have $<20 \%$ $r$-process contributions, and they could be used in the fitting process as $s$-process calibrating points.  An attempt to reproduce the two-component fitting with $s$-process only nuclei using the latest abundances and cross section values was not successful. Substitution of the current values with the older abundances \citep{89And} and cross sections \citep{87Bao} improves the situation; however, the overall agreement is not satisfactory. Further analysis explains this finding and suggests  a preference for the main $s$-process component fitting  because of several issues in the A$<$90 region
\begin{itemize}
\item Small number of $s$-process only medium nuclei ($^{70}$Ge,$^{76}$Se and $^{82}$Kr).
\item A limited number of measurements with these nuclei \citep{14Otu}.
\item Abundances of medium nuclei are not in equilibrium per the golden rules of $s$-process \citep{16Lug}.
\end{itemize}
Therefore, the main component ($N \geq$ 50 nuclei) is examined in the present work, and the list  of $s$-process only nuclei is based on Ref. \citep{82Kap}.  Neutron fluence parameters for $s$-process  only isotopes  were deduced using  Eq. \ref{eq:class3} above using the latest data for cross sections and solar system abundances.  Later, the deduced parameters were optimized using  least squares procedures, and  $f$ and $\tau_{0}$  neutron fluence distribution numerical values were obtained. The resulting fluence parameters  are shown in Table \ref{tab:sfits}.

\begin{table}[h!]
\begin{center}
 \caption{$S$-process strong component neutron fluence parameters for ENDF/B-VIII.0 and TENDL-2015 libraries \citep{18Bro,15Kon} cross sections and solar system abundances \citep{09Lod}.\label{tab:sfits}}
 \begin{tabular}{ >{\small}m{2.5cm} | >{\small}c | >{\small}c}
     \hline\hline
Parameters	&	ENDF/B-VIII.0	&	TENDL-2015	\\ \hline
f		&	0.000402$\pm$0.00083	&	0.000334$\pm$0.000052		\\
$\tau_{0}$	&	0.332365$\pm$0.027410	&	0.400150$\pm$0.034990		\\    \hline\hline
  \end{tabular}
\end{center}
\end{table}

\subsection{Computed values for s- and r-process abundances}

\begin{figure}
\begin{center}
\includegraphics[width=0.4\textwidth]{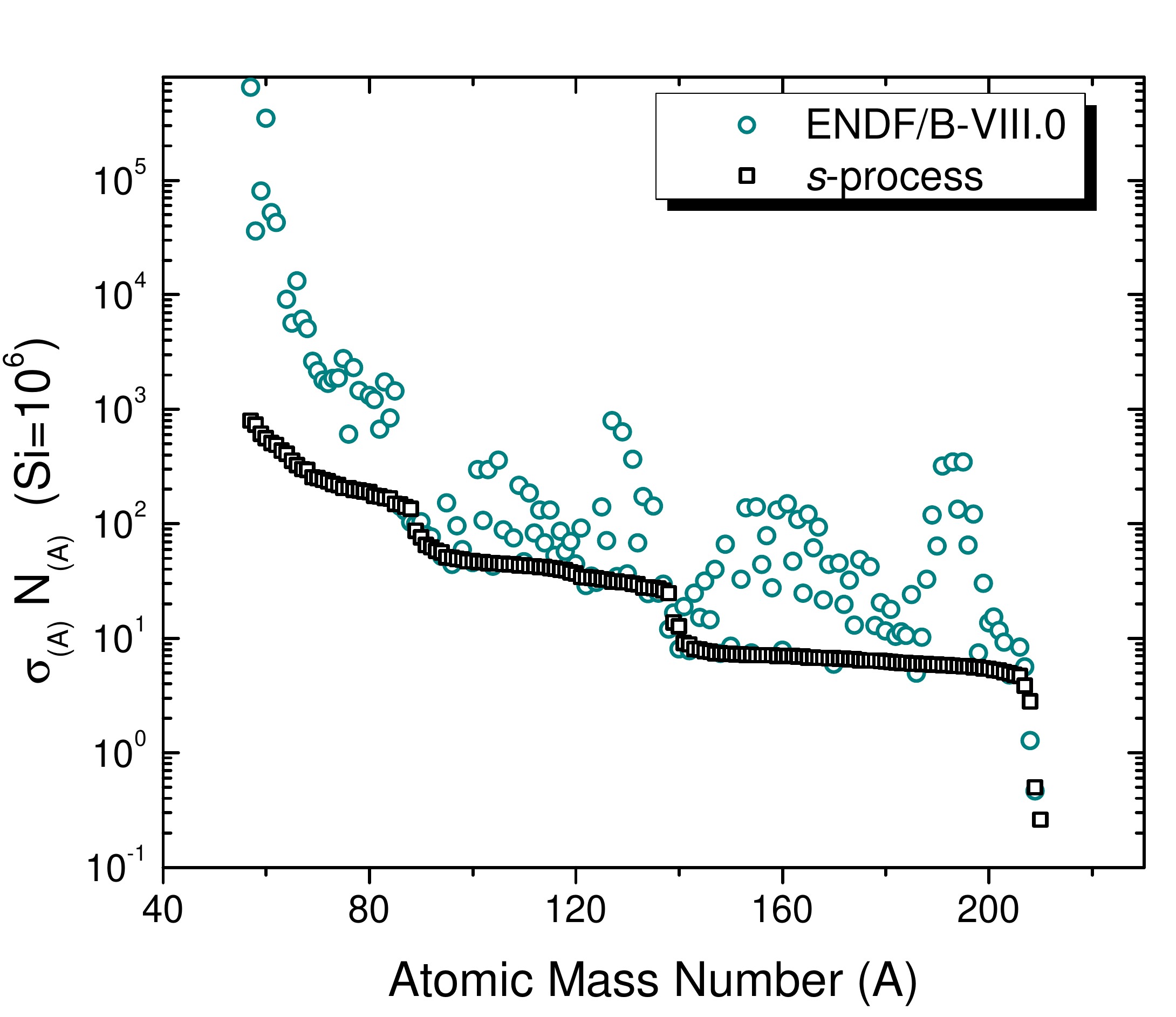}
\end{center}
\caption{Expected  and evaluated $s$-process MACS at $kT$=30 keV  abundances product values for  ENDF/B-VIII.0 nuclei. The solar system abundances are taken from Ref. \citep{09Lod}. \label{fig:snuclei}}
\end{figure}

$S$-process contribution to solar system abundances can be estimated and compared with 
 observed values using  neutron fluence parameters. The expected slow-neutron capture and ENDF/B-VIII.0 MACS ($kT$=30 keV) abundance product values are shown in Fig. \ref{fig:snuclei}. 
The Figure data indicate a surplus production for many nuclei compared with the $s$-process expectations. 
This surplus is commonly attributed to the $r$-process contribution, and it can be deduced  by subtracting the 
expected $s$-process production from the total values.

\begin{figure*}
\gridline{\fig{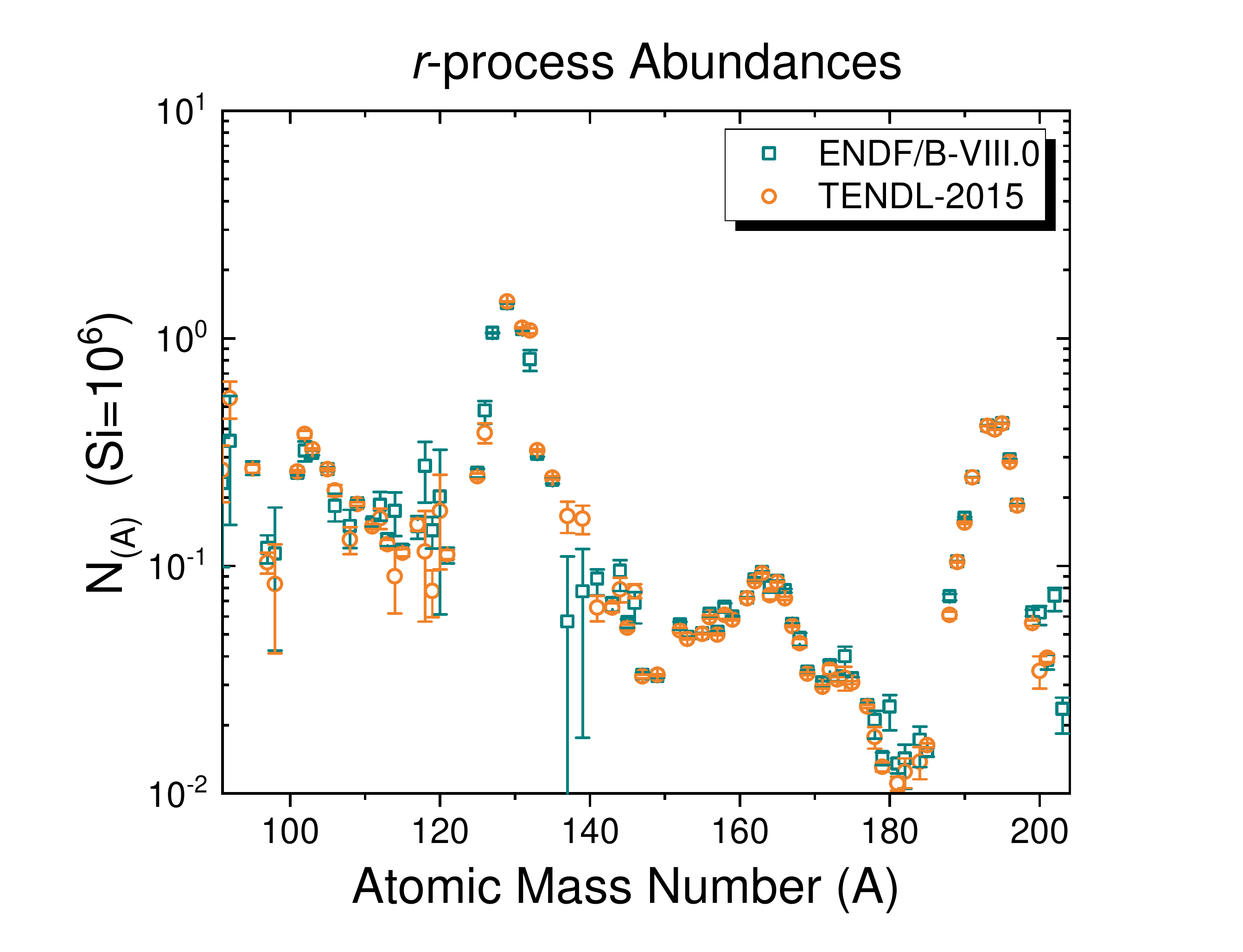}{0.4\textwidth}{(a)}
          }
\gridline{\fig{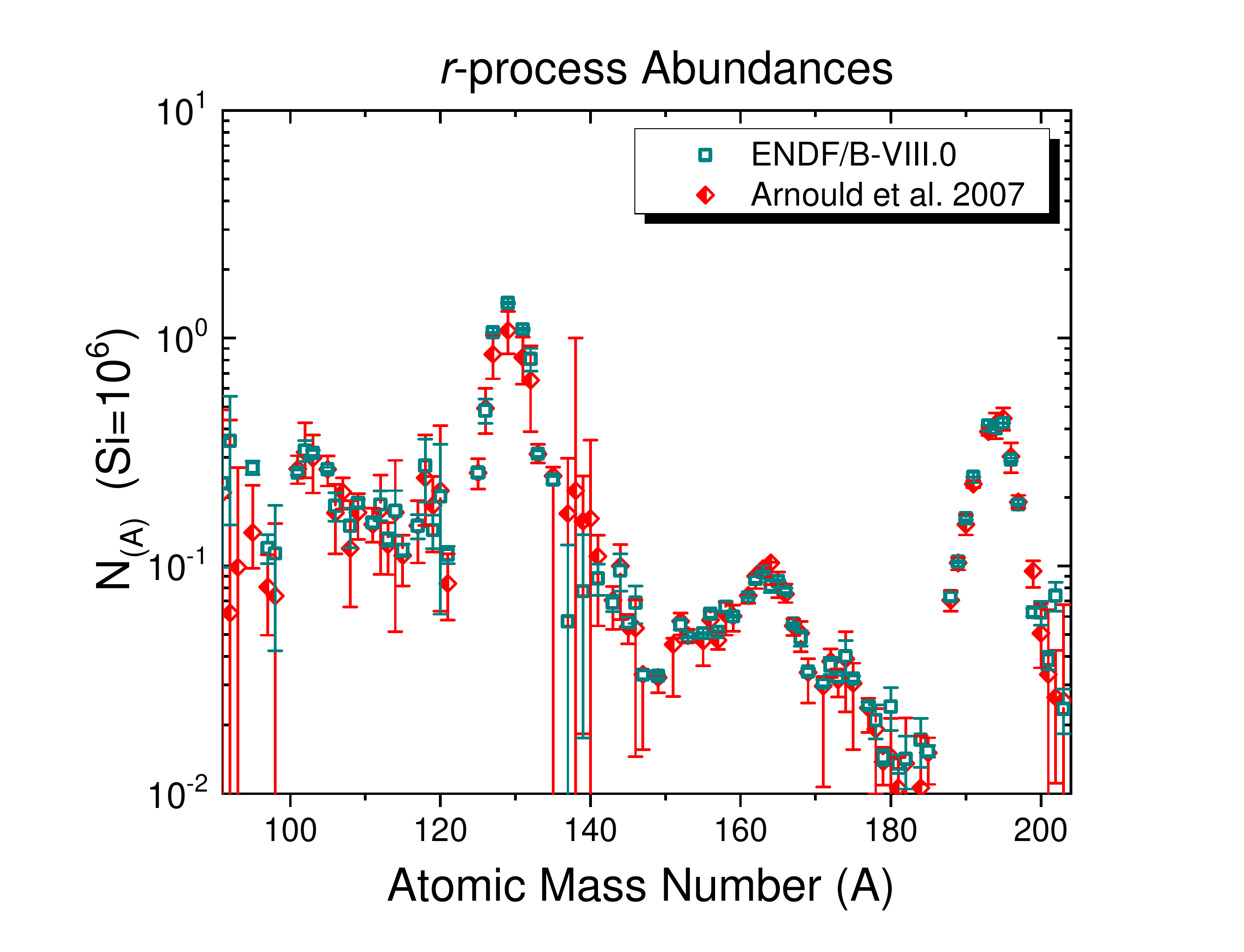}{0.4\textwidth}{(b)}
          }
\caption{Analysis of $r$-process abundances in evaluated nuclear data libraries:  Present work ENDF/B-VIII.0 and TENDL-2015 abundances (a) and comparison of  ENDF/B-VIII.0 and  solar system abundances \citep{93Pal,07Arn}  (b) .} \label{fig:rnuclei}
\end{figure*}

ENDF/B-VIII.0 and TENDL-2015 libraries $r$-process abundances  are shown in Fig. \ref {fig:rnuclei}, and their 
numerical values  are given in Table \ref{rprocessTable}. The table $r$-process only nuclei abundances are taken from Ref. \citep{09Lod}  for complementary comparison with   the solar system $r$-process abundances   \citep{93Pal,07Arn}. A Fig. \ref{fig:rnuclei} A data analysis shows two strong $r$-process abundance peaks and the broad surge due to production of lanthanides that were tentatively observed in neutron stars merger \citep{17Pia,17Sma,17Tan}.  These findings provide supplementary data for analysis of  star mergers as  potential  $r$-process astrophysical sites and highlights the importance of nuclear data libraries.  The  current paper contains the final results on ENDF libraries $r$-process abundances, while the preliminary results were reported elsewhere \citep{19Pri}. 
\include{rprocesstable20}

\section{Several applications of s-process nucleosynthesis calculations}

There are several avenues for utilization of the current work findings in  fundamental and applied sciences.  Here I will consider  $r$-process abundances, nuclear data verification and deduction of capture cross section benchmark values  for $A$=90-204 nuclei.

\subsection{Analysis of the  $R$-process abundances}

Examination of the  $r$-process abundances shown in  Table \ref{rprocessTable} indicate the ENDF/B-VIII.0 and TENDL-2015 values concur with solar system abundances that were obtained by Arnould et al. \citep{07Arn} from Ref. \citep{93Pal} with an exception of $^{88}$Sr $N$=50 and $N$=82 $^{138}$Ba and $^{140}$Ce lanthanide nuclei where the current work data are not smooth and show structure. This finding is consistent with  Kratz et al. who explored  $r$-processes by comparing calculations and observations in low-metallicity stars: ``The resulting $r$-process abundances may be very accurate for those isotopes with little $s$-process contribution (e.g.,$^{151}$Eu and $^{153}$Eu), but have significant uncertainties for isotopes where the $s$-process fraction is dominant (e.g., $^{138}$Ba, the most abundant of the seven naturally occurring Ba isotopes)." \citep{07Kra}. Indeed subtraction of $s$-process contribution from the total may result in  negative abundances including (1-N/N$_s$)=-8\% value for $^{138}$Ba \citep{93Pal}. This ``unnatural" finding was later interpreted by Arnould et al. as a positive $r$-process abundance of 0.214$^{+0.786}_{-0.214}$ (Si=10$^6$) \citep{07Arn}.

The aforementioned discussion on issues with subtraction of comparable values is not unique to nuclear astrophysics. Many EXFOR library  data entries \citep{14Otu} contain multiple negative cross section data points that were produced after background subtractions such as in double differential residual cross sections of $^{235}$U \citep{75Go}.  The similar effect was also noticed in $\beta\beta$-decay searches when samples with natural abundance were subtracted from  enriched samples. For example, the discoveries of $\beta\beta (2 \nu)$-decay in $^{76}$Ge and $^{100}$Mo include final spectra with negative data points \citep{90Va,90Vas}. Consequently, we have to conclude that subtraction of inexact nuclear data sets from each other may lead to unexpected results, and these results should be discussed and the observed imperfections  archived.  

Further analysis of  Table \ref{rprocessTable} data indicate negative TENDL-2015 abundances for $^{127}$I, $^{202}$Hg and $^{203}$Tl.   In this case, it is safe to assume that these results originate from the TENDL-2015 library deficiencies. The detailed discussion of these deviations is beyond the scope of the present paper, and these issues will provide additional motivation for the library improvements.   In addition, ENDF strontium, barium and cerium neutron cross sections are based on the SubGroup 23 (International library of fission product evaluations) recommendations \citep{SG23} that clearly have to be revisited.  All these cases should be further studied for either evaluated  cross sections \citep{18Bro,15Kon} or abundances deficiencies \citep{13Led,09Lod}, and additional neutron capture measurements for closed-neutron-shell nuclei are needed.  Finally, in spite of multiple differences between data sources and models, the overall agreement between the evaluated nuclear data libraries and  solar system abundances is very good.

\subsection{Astrophysical Benchmarks for Neutron Capture Cross Sections}

The systematics of evaluated cross sections combined with solar system abundances are very useful for analysis of data quality in evaluated libraries.  During the stellar nucleosynthesis modeling fitting procedure,  several   classical model outliers have been observed. The classical model fitting indicates that $^{116}$Sn $\sigma (A) N(A)$  product value is problematic in many libraries. The similar issues are observed for tellurium nuclei, while limitations in experimental data contribute to the issues with $^{76}$Se in ENDF/B-VIII.0
\begin{itemize} 
\item ENDF/B-VIII.0: $^{76}$Se, $^{116}$Sn, $^{122}$Te(n,$\gamma$).
\item TENDL-2015: $^{116}$Sn, $^{123,124}$Te(n,$\gamma$). 
\end{itemize}

These  potential data deficiencies provide directions for future measurements and evaluated libraries data improvements.  Furthermore, the $s$-process fitting results have  implications on  the neutron cross section measurements of uranium and plutonium fission products.  From the current results, it is easy to deduce the lower limit on Maxwellian-averaged cross sections by analyzing the calculated $s$-process product values and presently-observed abundances. Thus, ignoring the possible $r$-process contributions,  the lower bounds on neutron capture cross sections in the fission product region can be obtained. These cross sections  and complementary   ENDF library fission products analysis will be published separately.

\section{Conclusions \& Outlook}

The indirect findings of $r$-process elements in the neutron stars merger renewed interest in stellar nucleosynthesis 
calculations and the corresponding nuclear data. 
Recent release of the ENDF/B-VIII.0 evaluated nuclear data library  creates a unique opportunity for nuclear science and technology developments, and it has to be explored in nuclear astrophysics.

The Maxwellian-averaged cross sections and astrophysical reaction rates for 557 ENDF/B-VIII.0 and 2809 TENDL-2015 libraries target nuclides have been produced. These data were combined with the solar system abundances and fitted. Astrophysical $r$-process 
abundances have been extracted in the present work and compared with  available values. These large scale calculations of MACS and reaction rates open  new horizons for stellar nucleosynthesis calculations and other potential applications.

The present work demonstrates a large nuclear astrophysics potential of evaluated libraries 
and mutually dependent relations between nuclear industry and science data efforts. Further analysis reveals  potential data deficiencies, a strong need for 
new cross section measurements, and further clarifications of solar system and interstellar abundances. 
The next stage of the current project will involve incorporation of the evaluated nuclear data libraries into astrophysical model codes. 
Work on ENDF/B-VIII.0 $\&$ TENDL-2015 libraries data transfer for (n,$\gamma$), (n,$\alpha$), (n,p) and (n,fission) reaction channels within 0.01-10 GK neutron temperatures into REACLIB formatted library  \citep{00Rau,01Rau,10Cyb} is currently underway. The new data sets will provide a more extensive coverage of atomic nuclei and will make REACLIB fits more reliable across the whole $s$-process temperature range of 8-90 keV \citep{11Kap}. 

\section{Acknowledgments}
\label{sec:Acknowledgements}
The author is indebted to Dr. A. Sonzogni for support of this project, Dr. D. Brown for useful comments and Dr. V. Unferth  for a careful reading of the manuscript and valuable suggestions.  
Work at Brookhaven was funded by the Office of Nuclear Physics, Office of Science of the U.S. Department
of Energy, under Contract No. DE-AC02-98CH10886 with Brookhaven Science Associates, LLC.

\appendix

\section{ENDF/B-VIII.0, TENDL-2015 and KADoNIS 0.3 Maxwellian-averaged Cross Sections and Astrophysical Reaction Rates}
The present work MACS and astrophysical reaction rates are shown in Table \ref {tableCSRR}.
\include{AstroTableCSRR}

\end{document}

%% file: rprocesstable20.tex

\begin{small}

\begin{longtable}[h!]{@{\extracolsep\fill}lp{3.0cm}p{3.0cm}p{3.0cm}p{3.0cm}@{}}
\caption{{\it R}-process abundances obtained from ENDF/B-VIII.0 and  TENDL-2015 evaluated nuclear data libraries \citep{18Bro,15Kon} vs. solar system values \citep{93Pal,07Arn}. $R$-process only abundances for ENDF/TENDL libraries are adopted from 4.56 Gy ago values of Ref. \citep{09Lod}; 
alpha, beta and double-beta decay half lives (T$_{1/2}$) are gathered from the Evaluated Nuclear Structure Data File \citep{ensdf} and Ref. \citep{14Pri}, respectively. 
} \label{rprocessTable} \\ 
\hline\hline \\
 {\bfseries Target} &    {\bfseries  ENDF/B-VIII.0} & {\bfseries TENDL-2015}  & {\bfseries Solar System} & {\bfseries Comments} \\

\hline \\
\endfirsthead
\caption[]{{\it R}-process abundances obtained from ...  (continued).} \\
\hline\hline \\
 {\bfseries Target} &    {\bfseries  ENDF/B-VIII.0} & {\bfseries TENDL-2015}  & {\bfseries Solar System} & {\bfseries Comments} \\
\hline \\
\endhead 

31-Ga-69		&	1.988E+1$_{-1.657E-1}^{+3.533E-1}$	&	2.045E+1$_{-8.410E-2}^{+1.231E-1}$	&	6.180E+0$_{-6.180E+0}^{+3.190E+0}$	&			\\
32-Ge-70		&			     			&			     			&		&		$s$-process only	\\
30-Zn-70		&		8.000E+0	     			&		8.000E+0	     			&	7.740E+0$_{-9.400E-1}^{+8.100E-1}$	&		$r$-process only, T$_{1/2}$ $\geq$ 2.3$\times$10$^{17}$ y	\\
31-Ga-71		&	1.266E+1$_{-2.163E-1}^{+3.200E-1}$	&	1.269E+1$_{-1.432E-1}^{+1.509E-1}$	&	1.960E+0$_{-1.960E+0}^{+7.650E+0}$	&			\\
32-Ge-72		&	2.733E+1$_{- 4.898E-1}^{+7.188E-1}$	&	2.894E+1$_{-2.074E-1}^{+2.182E-1}$	&	0.000E+0$_{-0.000E+0}^{+9.930E+0}$	&			\\
32-Ge-73		&	7.756E+0$_{-1.182E-1}^{+1.704E-1}$	&	8.062E+0$_{-5.550E-2}^{+5.824E-2}$	&	6.310E+0$_{-6.310E+0}^{+1.880E+0}$	&			\\
32-Ge-74		&	3.645E+1$_{-5.393E-1}^{+7.735E-1}$	&	3.587E+1$_{-4.007E-1}^{+4.202E-1}$	&	1.970E+1$_{-9.760E+0}^{+9.200E+0}$	&			\\
33-As-75		&	5.650E+0$_{-5.160E-2}^{+7.249E-2}$	&	5.711E+0$_{-2.933E-2}^{+3.063E-2}$	&	3.780E+0$_{-5.400E-1}^{+9.000E-1}$	&			\\
34-Se-76		&			     			&			     			&		&		$s$-process only	\\
32-Ge-76		&		8.500E+0	     			&		8.500E+0	     			&	8.780E+0$_{-9.400E-1}^{+9.000E-1}$	&		$r$-process only	\\
34-Se-77		&	4.712E+0$_{-5.062E-2}^{+7.024E-2}$	&	4.809E+0$_{-2.569E-2}^{+2.679E-2}$	&	3.760E+0$_{-2.800E-1}^{+8.900E-1}$	&			\\
34-Se-78		&	1.390E+1$_{-2.471E-1}^{+3.421E-1}$	&	1.349E+1$_{-1.924E-1}^{+2.006E-1}$	&	0.000E+0$_{-0.000E+0}^{+1.030E+1}$	&			\\
34-Se-79		&			     			&			     			&	4.810E+0$_{-3.892E+0}^{+9.000E-1}$	&		$s$-process branching, T$_{1/2}$ = 3.26$\times$10$^{5}$ y	\\
34-Se-80		&	2.875E+1$_{-5.509E-1}^{+7.529E-1}$ 	&	3.001E+1$_{-2.624E-1}^{+2.728E-1}$	&	2.810E+1$_{-3.300E+0}^{+4.100E+0}$	&		$s$-process branching	\\
35-Br-81		&	4.524E+0$_{-8.912E-2}^{+1.190E-1}$	&	4.689E+0$_{-4.473E-2}^{+4.636E-2}$	&	4.070E+0$_{-1.030E+0}^{+8.000E-1}$	&		$s$-process branching	\\
34-Se-82		&		5.890E+0	     			&		5.890E+0	     			&	6.200E+0$_{-3.700E-1}^{+3.100E-1}$	&		$r$-process only, T$_{1/2}$ = 9.2$\times$10$^{19}$ y	\\
36-Kr-82		&			     			&			     			&		&		$s$-process only	\\
36-Kr-83		&	5.829E+0$_{-7.370E-2}^{+9.710E-2}$	&	6.071E+0$_{-2.871E-2}^{+2.969E-2}$	&	4.380E+0$_{-1.330E+0}^{+1.300E+0}$	&			\\
36-Kr-84		&	2.557E+1$_{-7.387E-1}^{+9.696E-1}$ 	&	2.755E+1$_{-3.203E-1}^{+3.311E-1}$	&	2.360E+1$_{-9.400E+0}^{+1.090E+1}$	&			\\
37-Rb-85		&	4.599E+0$_{-6.315E-2}^{+8.021E-2}$ 	&	4.638E+0$_{-3.668E-2}^{+3.777E-2}$	&	2.870E+0$_{-1.820E+0}^{+1.140E+0}$	&		$s$-process branching	\\
38-Sr-86		&			     			&			     			&		&		$s$-process branching	\\
38-Sr-87		&			     			&			     			&		&		$s$-process branching	\\
38-Sr-88		&	-5.704E+0$_{-3.059E+0}^{+3.771E+0}$ 	&	-1.790E+0$_{-1.598E+0}^{+1.639E+0}$	&	4.090E+0$_{-4.090E+0}^{+6.600E-1}$	&		$s$-process mostly, N=50	\\
39-Y-89			&	6.187E-1$_{-5.197E-1}^{+5.746E-1}$	&	1.310E+0$_{-2.548E-1}^{+2.578E-1}$	&	1.110E+0$_{-1.110E+0}^{+7.000E-1}$	&			\\
40-Zr-90		&	1.530E+0$_{-5.295E-1}^{+5.641E-1}$	&	1.944E+0$_{-2.770E-1}^{+2.792E-1}$	&	2.610E+0$_{-1.350E+0}^{+4.000E-1}$	&			\\
40-Zr-91		&	2.304E-1$_{-1.316E-1}^{+1.346E-1}$	&	2.637E-1$_{-7.296E-2}^{+7.322E-2}$	&	2.100E-1$_{-2.100E-1}^{+2.630E-1}$	&			\\
40-Zr-92		&	3.536E-1$_{-2.020E-1}^{+2.039E-1}$	&	5.448E-1$_{-1.006E-1}^{+1.008E-1}$	&	6.200E-2$_{-6.200E-2}^{+3.750E-1}$	&			\\
41-Nb-93		&			     			&			     			&	9.870E-2$_{-9.870E-2}^{+1.713E-1}$	&		$s$-process branching, $^{93}$Zr T$_{1/2}$ = 1.61$\times$10$^{6}$ y	\\
40-Zr-94		&			     			&		&	0.000E+0$_{-0.000E+0}^{+6.020E-2}$	&		$s$-process branching	\\
42-Mo-95		&	2.695E-1$_{-1.865E-2}^{+1.779E-2}$ 	&	2.675E-1$_{-1.056E-2}^{+1.053E-2}$	&	1.400E-1$_{-4.240E-2}^{+8.600E-2}$	&			\\
42-Mo-96		&			     			&			     			&		&		$s$-process only	\\
40-Zr-96		&		3.020E-1	     			&		3.020E-1	     			&	0.000E+0$_{-0.000E+0}^{+2.500E-2}$	&		$r$-process only, T$_{1/2}$ = 2.35$\times$10$^{19}$ y	\\
42-Mo-97		&	1.197E-1$_{-1.746E-2}^{+1.648E-2}$	&	1.034E-1$_{-1.096E-2}^{+1.092E-2}$	&	8.080E-2$_{-3.120E-2}^{+3.120E-2}$	&			\\
42-Mo-98		&	1.134E-1$_{-7.091E-2}^{+6.678E-2}$	&	8.311E-2$_{-4.174E-2}^{+4.157E-2}$	&	7.390E-2$_{-7.390E-2}^{+7.910E-2}$	&			\\
43-Tc-99		&			     			&			     			&		&		T$_{1/2}$=2.111$\times$10$^{5}$ y	\\
44-Ru-100		&			     			&			     			&		&		$s$-process only	\\
42-Mo-100		&		2.500E-1	     			&		2.500E-1	     			&	2.260E-1$_{-1.600E-2}^{+2.400E-2}$	&		$r$-process only, T$_{1/2}$ = 7.3$\times$10$^{18}$ y	\\
44-Ru-101		&	2.566E-1$_{-6.646E-3}^{+6.173E-3}$ 	&	2.598E-1$_{-3.426E-3}^{+3.407E-3}$	&	2.670E-1$_{-3.700E-2}^{+3.800E-2}$	&			\\
44-Ru-102		&	3.208E-1$_{-3.384E-2}^{+3.140E-2}$	&	3.789E-1$_{-1.419E-2}^{+1.411E-2}$	&	3.150E-1$_{-7.100E-2}^{+1.100E-1}$	&			\\
45-Rh-103		&	3.136E-1$_{-7.935E-3}^{+7.331E-3}$	&	3.251E-1$_{-3.481E-3}^{+3.461E-3}$	&	2.970E-1$_{-8.800E-2}^{+7.800E-2}$	&			\\
46-Pd-104		&			     			&			     			&		&		$s$-process only	\\
44-Ru-104		&		3.320E-1	     			&		3.320E-1	     			&	3.370E-1$_{-3.900E-2}^{+4.600E-2}$	&		$r$-process only	\\
46-Pd-105		&	2.656E-1$_{-5.297E-3}^{+4.874E-3}$ 	&	2.658E-1$_{-2.900E-3}^{+2.883E-3}$	&	2.660E-1$_{-4.200E-2}^{+3.700E-2}$	&			\\
46-Pd-106		&	1.834E-1$_{-2.643E-2}^{+2.430E-2}$	&	2.142E-1$_{-1.216E-2}^{+1.209E-2}$	&	1.710E-1$_{-5.800E-2}^{+5.700E-2}$	&			\\
46-Pd-107		&			     			&			     			&		&		T$_{1/2}$=6.5$\times$10$^{6}$ y	\\
46-Pd-108		&	1.495E-1$_{-2.957E-2}^{+2.707E-2}$	&	1.303E-1$_{-1.774E-2}^{+1.763E-2}$	&	1.190E-1$_{-5.300E-2}^{+7.400E-2}$	&		$s$-process branching	\\
47-Ag-109		&	1.886E-1$_{-6.708E-3}^{+6.117E-3}$	&	1.871E-1$_{-3.794E-3}^{+3.769E-3}$	&	1.720E-1$_{-4.100E-2}^{+3.500E-2}$	&			\\
48-Cd-110		&			     			&			     			&		&		$s$-process only	\\
46-Pd-110		&		1.590E-1	     			&		1.590E-1	     			&	1.560E-1$_{-2.000E-2}^{+1.800E-2}$	&		$r$-process only	\\
48-Cd-111		&	1.550E-1$_{-6.510E-3}^{+5.908E-3}$ 	&	1.492E-1$_{-4.017E-3}^{+3.989E-3}$	&	1.520E-1$_{-2.500E-2}^{+2.700E-2}$	&			\\
48-Cd-112		&	1.858E-1$_{-2.752E-2}^{+2.495E-2}$	&	1.618E-1$_{-1.693E-2}^{+1.681E-2}$	&	1.760E-1$_{-8.390E-2}^{+8.400E-2}$	&			\\
48-Cd-113		&	1.308E-1$_{-8.685E-3}^{+7.844E-3}$	&	1.246E-1$_{-5.234E-3}^{+5.195E-3}$	&	1.240E-1$_{-3.240E-2}^{+3.100E-2}$	&	 T$_{1/2}$ = 8.0$\times$10$^{15}$ y	\\
48-Cd-114		&	1.745E-1$_{-3.941E-2}^{+3.555E-2}$	&	9.001E-2$_{-2.810E-2}^{+2.789E-2}$	&	1.720E-1$_{-1.205E-1}^{+1.180E-1}$	&		T$_{1/2}$ $\geq$ 2.1$\times 10^{18}$ y	\\
49-In-115		&	1.172E-1$_{-7.518E-3}^{+6.744E-3}$	&	1.142E-1$_{-4.330E-3}^{+4.295E-3}$	&	1.110E-1$_{-2.940E-2}^{+2.500E-2}$	&		T$_{1/2}$ =4 .41$\times$10$^{14}$ y 	\\
50-Sn-116		&			     			&			     			&		&		$s$-process only	\\
48-Cd-116		&		1.180E-1	     			&		1.180E-1	     			&	9.550E-2$_{-2.550E-2}^{+3.150E-2}$	&		$r$-process only, T$_{1/2}$ = 3.0$\times$10$^{19}$ y	\\
50-Sn-117		&	1.495E-1$_{-1.822E-2}^{+1.619E-2}$ 	&	1.517E-1$_{-9.731E-3}^{+9.645E-3}$	&	1.500E-1$_{-4.700E-2}^{+4.300E-2}$	&			\\
50-Sn-118		&	2.755E-1$_{-8.550E-2}^{+7.577E-2}$	&	1.157E-1$_{-5.884E-2}^{+5.831E-2}$	&	2.440E-1$_{-9.300E-2}^{+1.310E-1}$	&			\\
50-Sn-119		&	1.431E-1$_{-2.387E-2}^{+2.089E-2}$	&	7.747E-2$_{-1.800E-2}^{+1.782E-2}$	&	1.840E-1$_{-6.900E-2}^{+6.300E-2}$	&			\\
50-Sn-120		&	2.015E-1$_{-1.402E-1}^{+1.223E-1}$ 	&	1.743E-1$_{-7.782E-2}^{+7.701E-2}$	&	2.140E-1$_{-1.506E-1}^{+1.980E-1}$	&			\\
51-Sb-121		&	1.121E-1$_{-9.715E-3}^{+8.302E-3}$	&	1.122E-1$_{-5.203E-3}^{+5.142E-3}$	&	8.360E-2$_{-2.580E-2}^{+2.940E-2}$	&			\\
52-Te-122		&			     			&			     			&		&		$s$-process only	\\
50-Sn-122		&		1.670E-1	     			&		1.670E-1	     			&	1.520E-1$_{-1.520E-1}^{+2.800E-2}$	&		$r$-process only	\\
52-Te-123		&			     			&			     			&		&		$s$-process only, T$_{1/2}$$\geq$9.2$\times$10$^{16}$ y	\\
51-Sb-123		&		1.340E-1	     			&		1.340E-1	     			&	1.130E-1$_{-2.050E-2}^{+1.800E-2}$	&		$r$-process only	\\
52-Te-124		&			     			&			     			&		&		$s$-process only	\\
50-Sn-124		&		2.090E-1	     			&		2.090E-1	     			&	2.200E-1$_{-2.500E-2}^{+2.200E-2}$	&		$r$-process only, T$_{1/2}$ $\geq$ 1.2$\times$10$^{21}$ y	\\
52-Te-125		&	2.567E-1$_{-1.144E-2}^{+9.656E-3}$ 	&	2.477E-1$_{-6.798E-3}^{+6.712E-3}$	&	2.560E-1$_{-3.900E-2}^{+3.900E-2}$	&			\\
52-Te-126		&	4.812E-1$_{-5.960E-2}^{+5.022E-2}$	&	3.843E-1$_{-3.931E-2}^{+3.881E-2}$	&	4.920E-1$_{-1.100E-1}^{+1.090E-1}$	&			\\
53-I-127		&	1.057E+0$_{-6.351E-3}^{+5.297E-3}$	&	-1.141E+0$_{-1.746E-1}^{+1.723E-1}$	&	8.480E-1$_{-1.850E+0}^{+1.550E-1}$	&		TENDL is lower than experiments	\\
54-Xe-128		&			     			&			     			&		&			\\
52-Te-128		&		1.486E+0	     			&		1.486E+0			&	1.470E+0$_{-1.800E-1}^{+1.500E-1}$	&		$r$-process only, T$_{1/2}$ = 2.41$\times$10$^{24}$ y	\\
54-Xe-129		&	1.426E+0$_{-1.079E-2}^{+8.966E-3}$	&	1.450E+0$_{-3.789E-3}^{+3.728E-3}$	&	1.080E+0$_{-1.570E-1}^{+2.300E-1}$	&			\\
54-Xe-130		&			     			&			     			&		&			\\
52-Te-130		&		1.585E+0	     			&		1.585E+0	     			&	1.580E+0$_{-1.600E-1}^{+1.600E-1}$	&		$r$-process only, T$_{1/2}$ = 7.14$\times$10$^{20}$ y	\\
54-Xe-131		&	1.092E+0$_{-1.447E-2}^{+1.193E-2}$	&	1.112E+0$_{-6.122E-3}^{+6.023E-3}$	&	8.220E-1$_{-1.950E-1}^{+1.790E-1}$	&			\\
54-Xe-132		&	8.091E-1$_{-9.283E-2}^{+7.634E-2}$	&	1.080E+0$_{-2.795E-2}^{+2.749E-2}$	&	6.530E-1$_{-2.610E-1}^{+2.530E-1}$	&			\\
55-Cs-133		&	3.108E-1$_{-8.941E-3}^{+7.232E-3}$	&	3.215E-1$_{-3.870E-3}^{+3.805E-3}$	&	3.090E-1$_{-2.600E-2}^{+3.200E-2}$	&			\\
56-Ba-134		&			     			&			     			&		&			\\
54-Xe-134		&		5.270E-1	     			&		5.270E-1	     			&	3.850E-1$_{-1.540E-1}^{+9.200E-2}$	&		$r$-process only, T$_{1/2}$ $\geq$ 5.8$\times$10$^{22}$ y	\\
56-Ba-135		&	2.384E-1$_{-8.434E-3}^{+6.784E-3}$ 	&	2.438E-1$_{-3.999E-3}^{+3.931E-3}$	&	2.480E-1$_{-2.480E-1}^{+2.400E-2}$	&			\\
56-Ba-136		&			     			&			     			&		&			\\
54-Xe-136		&		4.290E-1	     			&		4.290E-1	     			&	3.300E-1$_{-7.000E-2}^{+6.600E-2}$	&		$r$-process only, T$_{1/2}$ = 2.34$\times$10$^{21}$ y	\\
56-Ba-137		&	5.703E-2$_{-6.661E-2}^{+5.289E-2}$	&	1.657E-1$_{-2.629E-2}^{+2.582E-2}$	&	1.700E-1$_{-1.700E-1}^{+1.260E-1}$	&			\\
56-Ba-138		&	-3.439E+0$_{-9.999E-1}^{+7.834E-1}$	&	-3.367E+0$_{-5.139E-1}^{+5.047E-1}$	&	2.140E-1$_{-2.140E-1}^{+7.860E-1}$	&		$s$-process mostly, N=82	\\
57-La-139		&	7.730E-2$_{-5.975E-2}^{+4.163E-2}$ 	&	1.610E-1$_{-2.320E-2}^{+2.276E-2}$	&	1.570E-1$_{-1.387E-1}^{+9.100E-2}$	&			\\
58-Ce-140		&	-5.995E-1$_{-2.604E-1}^{+1.778E-1}$	&	-4.663E-1$_{-1.183E-1}^{+1.161E-1}$	&	1.610E-1$_{-1.610E-1}^{+1.960E-1}$	&		$s$-process mostly, N=82	\\
59-Pr-141		&	8.802E-2$_{-1.366E-2}^{+8.675E-3}$	&	6.552E-2$_{-8.351E-3}^{+8.205E-3}$	&	1.100E-1$_{-5.550E-2}^{+2.600E-2}$	&			\\
60-Nd-142		&			     			&			     			&		&		$s$-process only	\\
58-Ce-142		&		1.310E-1	     			&		1.310E-1	     			&	6.600E-2$_{-6.600E-2}^{+6.500E-2}$	&		$r$-process only, T$_{1/2}$ $>$ 5.0$\times$10$^{16}$ y	\\
60-Nd-143		&	6.871E-2$_{-5.632E-3}^{+3.477E-3}$	&	6.559E-2$_{-2.934E-3}^{+2.885E-3}$	&	7.060E-2$_{-1.800E-2}^{+1.050E-2}$	&			\\
60-Nd-144		&	9.517E-2$_{-1.773E-2}^{+1.091E-2}$	&	7.888E-2$_{-9.735E-3}^{+9.572E-3}$	&	9.980E-2$_{-4.160E-2}^{+2.420E-2}$	&		T$_{1/2}$ = 2.29$\times$10$^{15}$ y	\\
60-Nd-145		&	5.653E-2$_{-3.047E-3}^{+1.856E-3}$	&	5.350E-2$_{-1.686E-3}^{+1.658E-3}$	&	5.400E-2$_{-8.400E-3}^{+7.100E-3}$	&			\\
60-Nd-146		&	6.883E-2$_{-1.290E-2}^{+7.846E-3}$	&	7.736E-2$_{-5.462E-3}^{+5.373E-3}$	&	5.330E-2$_{-3.880E-2}^{+1.780E-2}$	&			\\
62-Sm-147		&			4.1000E-2     			&		4.1000E-2  	     			&	3.340E-2$_{-1.780E-2}^{+1.300E-3}$	&		T$_{1/2}$=1.060$\times$10$^{11}$ y	\\
62-Sm-148		&			     			&			     			&		&		$s$-process only, T$_{1/2}$=7$\times$10$^{15}$ y	\\
60-Nd-148		&		4.900E-2	     			&		4.900E-2	     			&	4.210E-2$_{-2.000E-2}^{+1.010E-2}$	&		$r$-process only	\\
62-Sm-149		&	3.287E-2$_{-6.851E-4}^{+4.118E-4}$	&	3.317E-2$_{-3.002E-4}^{+2.954E-4}$	&	3.230E-2$_{-4.500E-3}^{+5.000E-4}$	&			\\
62-Sm-150		&			     			&			     			&		&			\\
60-Nd-150		&		4.800E-2	     			&		4.800E-2	     			&	4.900E-2$_{-3.100E-3}^{+2.500E-3}$	&		$r$-process only, T$_{1/2}$ = 0.91$\times$10$^{19}$ y	\\
62-Sm-151		&			     			&			     			&		&		T$_{1/2}$=90 y 	\\
62-Sm-152		&	5.511E-2$_{-2.636E-3}^{+1.580E-3}$	&	5.211E-2$_{-1.481E-3}^{+1.458E-3}$	&	5.710E-2$_{-2.300E-3}^{+4.100E-3}$	&		$s$-process branching	\\
63-Eu-153	 	&	4.868E-2$_{-4.511E-4}^{+2.700E-4}$	&	4.785E-2$_{-2.787E-4}^{+2.742E-4}$	&	4.950E-2$_{-3.500E-3}^{+3.100E-3}$	&		$s$-process branching	\\
64-Gd-154		&			     			&			     			&		&		$s$-process only	\\
62-Sm-154		&		6.000E-2	     			&		6.000E-2	     			&	5.950E-2$_{-9.000E-3}^{+1.400E-3}$	&		$r$-process only	\\
64-Gd-155		&	5.054E-2$_{-4.580E-4}^{+2.738E-4}$	&	5.025E-2$_{-2.395E-4}^{+2.356E-4}$	&	4.680E-2$_{-1.040E-2}^{+3.200E-3}$	&			\\
64-Gd-156		&	6.157E-2$_{-1.998E-3}^{+1.194E-3}$	&	5.956E-2$_{-1.101E-3}^{+1.084E-3}$	&	5.790E-2$_{-7.800E-3}^{+5.500E-3}$	&			\\
64-Gd-157		&	5.116E-2$_{-8.531E-4}^{+5.093E-4}$	&	4.996E-2$_{-4.975E-4}^{+4.896E-4}$	&	4.710E-2$_{-4.200E-3}^{+3.700E-3}$	&			\\
64-Gd-158		&	6.610E-2$_{-3.872E-3}^{+2.310E-3}$	&	6.083E-2$_{-2.241E-3}^{+2.205E-3}$	&	6.140E-2$_{-1.170E-2}^{+8.000E-3}$	&			\\
65-Tb-159		&	5.998E-2$_{-5.683E-4}^{+3.383E-4}$	&	5.812E-2$_{-4.142E-4}^{+4.077E-4}$	&	6.010E-2$_{-8.400E-3}^{+7.100E-3}$	&			\\
66-Dy-160		&			     			&			     			&		&		$s$-process only	\\
64-Gd-160		&		7.870E-2	     			&		7.870E-2	     			&	7.410E-2$_{-8.600E-3}^{+4.600E-3}$	&		$r$-process only, T$_{1/2}$ $\geq$ 3.1$\times$10$^{19}$ y	\\
66-Dy-161		&	7.258E-2$_{-6.016E-4}^{+3.577E-4}$ 	&	7.195E-2$_{-3.335E-4}^{+3.282E-4}$	&	7.410E-2$_{-5.700E-3}^{+4.000E-4}$	&			\\
66-Dy-162		&	8.729E-2$_{-2.580E-3}^{+1.533E-3}$	&	8.560E-2$_{-1.349E-3}^{+1.328E-3}$	&	9.000E-2$_{-1.050E-2}^{+1.700E-3}$	&			\\
66-Dy-163		&	9.404E-2$_{-1.075E-3}^{+6.382E-4}$	&	9.300E-2$_{-5.884E-4}^{+5.792E-4}$	&	9.720E-2$_{-8.200E-3}^{+8.000E-4}$	&			\\
66-Dy-164		&	8.187E-2$_{-5.366E-3}^{+3.182E-3}$	&	7.434E-2$_{-3.118E-3}^{+3.069E-3}$	&	1.030E-1$_{-2.030E-2}^{+1.000E-3}$	&			\\
67-Ho-165		&	8.580E-2$_{-8.663E-4}^{+5.122E-4}$	&	8.497E-2$_{-4.726E-4}^{+4.652E-4}$	&	8.390E-2$_{-1.110E-2}^{+1.020E-2}$	&			\\
68-Er-166		&	7.812E-2$_{-1.646E-3}^{+9.727E-4}$	&	7.202E-2$_{-1.253E-3}^{+1.234E-3}$	&	7.530E-2$_{-6.200E-3}^{+8.000E-3}$	&			\\
68-Er-167		&	5.558E-2$_{-7.364E-4}^{+4.347E-4}$	&	5.434E-2$_{-4.440E-4}^{+4.371E-4}$	&	5.460E-2$_{-5.100E-3}^{+4.000E-3}$	&			\\
68-Er-168		&	4.833E-2$_{-3.780E-3}^{+2.230E-3}$	&	4.582E-2$_{-1.975E-3}^{+1.945E-3}$	&	5.060E-2$_{-8.600E-3}^{+6.400E-3}$	&			\\
69-Tm-169		&	3.430E-2$_{-1.051E-3}^{+6.185E-4}$	&	3.361E-2$_{-5.481E-4}^{+5.396E-4}$	&	3.400E-2$_{-9.000E-3}^{+5.100E-3}$	&			\\
70-Yb-170		&			     			&			     			&		&		$s$-process only	\\
68-Er-170		&		3.900E-2	     			&		3.900E-2	     			&	3.690E-2$_{-8.600E-3}^{+3.800E-3}$	&		$r$-process only	\\
70-Yb-171		&	3.072E-2$_{-8.989E-4}^{+5.285E-4}$	&	2.945E-2$_{-5.214E-4}^{+5.134E-4}$	&	2.970E-2$_{-1.800E-2}^{+2.900E-3}$	&			\\
70-Yb-172		&	3.657E-2$_{-3.177E-3}^{+1.867E-3}$	&	3.512E-2$_{-1.606E-3}^{+1.581E-3}$	&	3.810E-2$_{-5.800E-3}^{+5.100E-3}$	&			\\
70-Yb-173		&	3.264E-2$_{-1.446E-3}^{+8.483E-4}$	&	3.174E-2$_{-7.501E-4}^{+7.386E-4}$	&	3.160E-2$_{-5.000E-3}^{+3.700E-3}$	&			\\
70-Yb-174		&	4.004E-2$_{-7.030E-3}^{+4.119E-3}$	&	3.221E-2$_{-3.913E-3}^{+3.853E-3}$	&	3.910E-2$_{-1.620E-2}^{+2.400E-3}$	&			\\
71-Lu-175		&	3.201E-2$_{-8.344E-4}^{+4.869E-4}$	&	3.071E-2$_{-4.934E-4}^{+4.859E-4}$	&	3.050E-2$_{-1.490E-2}^{+6.900E-3}$	&			\\
71-Lu-176		&			     			&			     			&		&		$s$-process only, T$_{1/2}$=3.76$\times$10$^{10}$ y	\\
70-Yb-176		&		3.330E-2	     			&		3.330E-2	     			&	2.920E-2$_{-1.150E-2}^{+4.200E-3}$	&		$r$-process only	\\
72-Hf-177		&	2.451E-2$_{-7.513E-4}^{+4.379E-4}$	&	2.412E-2$_{-3.831E-4}^{+3.773E-4}$	&	2.380E-2$_{-4.200E-3}^{+2.500E-3}$	&			\\
72-Hf-178		&	2.102E-2$_{-3.597E-3}^{+2.095E-3}$	&	1.767E-2$_{-1.947E-3}^{+1.918E-3}$	&	1.920E-2$_{-9.200E-3}^{+4.400E-3}$	&			\\
72-Hf-179		&	1.446E-2$_{-1.129E-3}^{+6.562E-4}$	&	1.311E-2$_{-6.342E-4}^{+6.248E-4}$	&	1.380E-2$_{-2.900E-3}^{+2.200E-3}$	&			\\
72-Hf-180		&	2.411E-2$_{-5.128E-3}^{+2.978E-3}$	&	-6.330E-4$_{-4.340E-3}^{+4.275E-3}$	&	1.450E-2$_{-1.450E-2}^{+6.900E-3}$	&	TENDL is lower than experiments		\\
73-Ta-181		&	1.353E-2$_{-1.254E-3}^{+7.260E-4}$	&	1.106E-2$_{-7.798E-4}^{+7.684E-4}$	&	1.060E-2$_{-6.400E-3}^{+3.500E-3}$	&			\\
74-W-182		&	1.419E-2$_{-3.711E-3}^{+2.146E-3}$	&	1.244E-2$_{-1.871E-3}^{+1.844E-3}$	&	1.360E-2$_{-1.360E-2}^{+7.900E-3}$	&			\\
74-W-183		&	8.727E-3$_{-1.827E-3}^{+1.054E-3}$	&	6.465E-3$_{-1.030E-3}^{+1.015E-3}$	&	6.500E-3$_{-6.500E-3}^{+3.500E-3}$	&		T$_{1/2}$$\geq$6.7$\times$10$^{20}$ y	\\
74-W-184		&	1.723E-2$_{-4.162E-3}^{+2.398E-3}$ 	&	1.378E-2$_{-2.213E-3}^{+2.181E-3}$	&	1.060E-2$_{-1.060E-2}^{+7.300E-3}$	&			\\
75-Re-185		&	1.537E-2$_{-8.971E-4}^{+5.154E-4}$	&	1.628E-2$_{-3.463E-4}^{+3.413E-4}$	&	1.510E-2$_{-4.100E-3}^{+2.500E-3}$	&			\\
76-Os-186		&			     			&			     			&		&		$s$-process only, T$_{1/2}$=2.0$\times$10$^{15}$ y	\\
74-W-186		&		3.900E-2	     			&		3.900E-2	     			&	2.450E-2$_{-1.720E-2}^{+9.200E-3}$	&		$r$-process only, T$_{1/2}$$\geq$2.3$\times$10$^{19}$ y	\\
76-Os-187		&			     			&			     			&		&		$s$-process only	\\
75-Re-187		&		3.740E-2	     			&		3.740E-2	     			&	3.180E-2$_{-4.800E-3}^{+4.100E-3}$	&		$r$-process only, T$_{1/2}$=4.33$\times$10$^{10}$ y	\\
76-Os-188		&	7.356E-2$_{-2.835E-3}^{+1.624E-3}$	&	6.085E-2$_{-2.317E-3}^{+2.284E-3}$	&	7.080E-2$_{-7.500E-3}^{+7.300E-3}$	&			\\
76-Os-189		&	1.044E-1$_{-9.420E-4}^{+5.386E-4}$	&	1.041E-1$_{-4.597E-4}^{+4.532E-4}$	&	1.030E-1$_{-6.900E-3}^{+6.000E-3}$	&			\\
76-Os-190		&	1.621E-1$_{-2.852E-3}^{+1.630E-3}$	&	1.552E-1$_{-1.868E-3}^{+1.841E-3}$	&	1.520E-1$_{-1.500E-2}^{+1.600E-2}$	&			\\
77-Ir-191		&	2.453E-1$_{-7.945E-4}^{+4.530E-4}$	&	2.446E-1$_{-4.271E-4}^{+4.211E-4}$	&	2.290E-1$_{-8.000E-3}^{+8.000E-3}$	&			\\
78-Pt-192		&			     			&			     			&		&		$s$-process only	\\
76-Os-192		&		2.780E-1	     			&		2.780E-1	     			&	2.730E-1$_{-2.100E-2}^{+1.600E-2}$	&		$r$-process only	\\
77-Ir-193		&	4.137E-1$_{-1.225E-3}^{+6.970E-4}$	&	4.121E-1$_{-6.962E-4}^{+6.865E-4}$	&	3.880E-1$_{-1.400E-2}^{+1.400E-2}$	&		$s$-process branching	\\
78-Pt-194		&	4.014E-1$_{-3.135E-3}^{+1.782E-3}$	&	3.973E-1$_{-1.781E-3}^{+1.757E-3}$	&	4.210E-1$_{-5.900E-2}^{+4.900E-2}$	&			\\
78-Pt-195		&	4.237E-1$_{-1.236E-3}^{+7.005E-4}$ 	&	4.220E-1$_{-7.026E-4}^{+6.930E-4}$	&	4.450E-1$_{-5.100E-2}^{+4.800E-2}$	&			\\
78-Pt-196		&	2.929E-1$_{-4.914E-3}^{+2.784E-3}$	&	2.864E-1$_{-2.795E-3}^{+2.757E-3}$	&	3.020E-1$_{-4.600E-2}^{+4.500E-2}$	&			\\
79-Au-197		&	1.857E-1$_{-1.574E-3}^{+8.884E-4}$ 	&	1.838E-1$_{-8.799E-4}^{+8.680E-4}$	&	1.910E-1$_{-1.200E-2}^{+1.300E-2}$	&			\\
80-Hg-198		&			     			&			     			&		&		$s$-process only	\\
78-Pt-198		&		9.100E-2	     			&		9.100E-2	     			&	9.500E-2$_{-1.450E-2}^{+1.000E-2}$	&		$r$-process only	\\
80-Hg-199		&	6.263E-2$_{-2.435E-3}^{+1.366E-3}$	&	5.609E-2$_{-1.639E-3}^{+1.618E-3}$	&	5.070E-2$_{-1.500E-2}^{+1.750E-2}$	&			\\
80-Hg-200		&	6.233E-2$_{-7.407E-3}^{+4.147E-3}$	&	3.454E-2$_{-5.603E-3}^{+5.529E-3}$	&	3.340E-2$_{-2.730E-2}^{+3.060E-2}$	&			\\
80-Hg-201		&	3.868E-2$_{-3.623E-3}^{+2.017E-3}$	&	3.945E-2$_{-1.611E-3}^{+1.591E-3}$	&	2.650E-2$_{-1.540E-2}^{+1.610E-2}$	&			\\
80-Hg-202		&	7.395E-2$_{-1.072E-2}^{+5.951E-3}$	&	-8.039E-2$_{-1.704E-2}^{+1.683E-2}$	&	2.570E-2$_{-2.570E-2}^{+4.100E-2}$	&		TENDL is lower than experiments	\\
81-Tl-203		&	2.352E-2$_{-5.199E-3}^{+2.860E-3}$	&	-1.124E-2$_{-5.114E-3}^{+5.056E-3}$	&	3.300E-3$_{-3.300E-3}^{+2.380E-2}$	&		TENDL is lower than experiments	\\
82-Pb-204		&			     			&			     			&		&		$s$-process only, T$_{1/2}$$\geq$1.4$\times$10$^{17}$ y	\\
80-Hg-204		&		3.100E-2	     			&		3.100E-2	     			&	2.660E-2$_{-9.500E-3}^{+6.400E-3}$	&		$r$-process only	\\
82-Pb-205		&			     			&			     			&		&		T$_{1/2}$=1.73$\times$10$^{7}$ y	\\
82-Pb-206		&			     			&			     			&	1.970E-1$_{-1.560E-1}^{+1.820E-1}$	&		isotope recycling	\\
82-Pb-207		&			     			&			     			&	1.420E-1$_{-1.420E-1}^{+2.930E-1}$	&		isotope recycling	\\
82-Pb-208		&			     			&			     			&	3.000E-4$_{-3.000E-4}^{+1.780E+0}$	&		isotope recycling	\\
83-Bi-209		&			     			&			     			&	5.010E-2$_{-4.010E-2}^{+1.139E-1}$	&		isotope recycling	\\
84-Po-210		&			     			&			     			&		&		T$_{1/2}$=138.376 d	\\

\hline \hline
\end{longtable}
\end{small}
